\newcommand{\be}{\begin{equation}}\newcommand{\ee}{\end{equation}}
\newcommand{\bea}{\begin{eqnarray}}\newcommand{\eea}{\end{eqnarray}}

\newcommand{\D}[1]{{\cal D}^{#1}}
\documentstyle[12pt]{article}

\begin{document}
\thispagestyle{empty}
\renewcommand{\thefootnote}{\fnsymbol{footnote}}
\begin{center}
{\bf N=3 SUPERSYMMETRIC EXTENSION OF KdV EQUATION} \vspace{1.5cm} \\
STEFANO BELLUCCI \vspace{1cm} \\
{\it INFN--Laboratori Nazionali di Frascati} \\
{\it P.O. Box 13, I-00044 Frascati, Italy} \vspace{1cm}\\
and \vspace{1cm}\\
EVGENYI IVANOV and SERGEY KRIVONOS \vspace{1cm}\\
{\it JINR--Laboratory of Theoretical Physics} \\
{\it Dubna, Head Post Office, P.O. Box 79, 101 000 Moscow, Russia}
\vspace{1.5cm} \\
{ABSTRACT}
\end{center}
\begin{quotation}
{\tenrm
We construct a one-parameter family of N=3 supersymmetric extensions of
the KdV equation as a Hamiltonian flow on N=3 superconformal algebra and
argue that it is non-integrable for any choice of the parameter. Then
we propose a modified N=3 super KdV equation which possesses the
higher order conserved quantities and so is a candidate for an
integrable system. Upon reduction to N=2, it yields the recently
discussed ``would-be integrable'' version of the N=2 super KdV equation.
In the bosonic core it contains a coupled system of the KdV type
equation and a three-component generalization of the mKdV equation.
We give a Hamiltonian formulation of the new N=3 super KdV equation
as a flow on some contraction of the direct sum of two N=3 superconformal
algebras.} \vspace{0.5cm} \\
\end{quotation}
\begin{center}
Submitted to {\it J. Math. Phys.}
\end{center}
\vfill
\newpage\setcounter{page}1

\section{Introduction}
In recent years there has been an incredible growth of interest in studying
integrable KdV-type hierarchies and their supersymmetric extensions, mainly
due to the distinguished role these systems play in $2D$ (super)gravities
and the related matrix models${}^{1-8}$.

A remarkable feature of the KdV hierarchy
is its relation, via
the second Hamiltonian structure, to the Virasoro algebra${}^2$. This
provides a link between
the KdV hierarchy and $2D$ conformal field theories (and $2D$ gravity).
The mKdV hierarchy is related in the same way to the $U(1)$ Kac-Moody
algebra, the famous Miura map being recognized
as the Sugawara-Feigin-Fuchs representation for the Virasoro algebra.
Analogously, nonlinear $W$-algebras and their various generalizations define
the second Hamiltonian structures for generalized KdV hierarchies which thus
turn out to be relevant to $W$-gravities and proper generalizations of the
latter.
For instance, Zamolodchikov's $W_3$-algebra amounts to the second
Hamiltonian structure for the Boussinesq hierarchy${}^3$. An important
implication of these relationships is the possibility to construct new
integrable systems of the KdV type and their superextensions in a regular way,
starting with the structure relations of one or another infinite-dimensional
algebra or superalgebra.

With making use of this approach, in refs.${}^{3-8}$ $N=1$ and $N=2$
supersymmetric KdV equations with $N=1$ and $N=2$ superconformal
algebras as the second Hamiltonian structure have been found and their
integrability properties have been studied. It is of interest to treat in
the same context higher $N$ superextensions of KdV, by relating them to the
higher $N$ superconformal algebras. Some preliminary steps in this direction
for the $N=3$ and $N=4$ cases (however, without any discussion of the
integrability issues) have been made in${}^{9,10}$. In the present
paper we
report on the results of a more thorough study of the $N=3$ case.

Before displaying the main content of our paper let us briefly recall
the precise meaning of the aforementioned interrelation between the KdV and
super KdV systems on the one hand and Virasoro and super Virasoro algebras
on the other.

As was shown in${}^{2}$, the KdV equation
\begin{equation}\label{a1}
u_t=-u_{xxx}+6uu_x
\end{equation}
can be treated as a Hamiltonian system,
$$
u_t= \left\{ u,{\cal H}\right\} \quad ,
$$
with the Hamiltonian and the Poisson brackets defined by
\begin{equation}\label{a2}
{\cal H} = \frac{1}{2} \int dx \; u^2(x) \quad , \;\;\;
\{ u(x),u(y) \} = \left[ -\partial^3+
  4u\partial+2u_x \right] \delta (x-y) \quad .
\end{equation}
Just this property is referred to as the existence of the second Hamiltonian
structure for the KdV equation. For the Fourier modes of $u(x)$,
\begin{equation}\label{a5}
u(x)  =  {6\over c}\sum_n \mbox{exp}(-inx)L_n -{1\over 4} \quad,
\end{equation}
the Poisson brackets in (\ref{a2}) imply the structure relations of
the Virasoro algebra
\begin{equation}
i\left\{ L_n,L_m \right\} =  (n-m)L_{n+m}+{c\over 12}(n^3-n)\delta_{n+m,0}\;.
   \label{a4}
\end{equation}
So, from the formal point of view, the  definition (\ref{a2}) means that the
density of the KdV Hamiltonian $\cal H$ is the square of
a conformal stress-tensor $u(x)$ obeying the Virasoro algebra
(\ref{a2}), (\ref{a4}).
Note that the Hamiltonian in eq. (1.2) has dimension 3 and is unique,
i.e. it is the only Hamiltonian of such a dimension that can be built out of
the dimension 2 field $u(x)$.
The higher order conserved quantities of the KdV equation can be regarded as
the Hamiltonians which generate, through the Poisson brackets (\ref{a2}),
next equations from the KdV hierarchy.

The same idea was applied for constructing $N=1$ and
$N=2$ superextensions of the KdV equation [3-8].
They were related in an
analogous way, via the second Hamiltonian structure, to
$N=1$ and $N=2$
superconformal algebras. In the latter case,
starting from $N=2$ superconformal algebra in the form
\begin{equation}\label{a6}
\left\{ \Phi (X),\Phi (X')\right\}=\left[ {\cal D}_1{\cal D}_2
\partial +2\Phi\partial -\left({\cal D}_1\Phi\right){\cal D}_1
-\left({\cal D}_2\Phi\right){\cal D}_2+2\Phi_x \right]
\Delta(X-X')
\end{equation}
with
$$
X\equiv \left\{x,\theta_1,\theta_2\right\}\quad , \quad
{\cal D}_i=\theta_i \partial+\frac{\partial}{\partial \theta_i} \quad ,
$$
$$
\Delta (X-X')=(\theta_2-\theta'_2)(\theta_1-\theta'_1)\delta (x-x')
$$
and choosing the most general $N=2$ supersymmetric Hamiltonian
of dimension 3
\begin{equation}\label{a7}
{\cal H}=\frac{1}{2}\int dx d^2 \theta \left( \Phi{\cal D}_1{\cal D}_2\Phi+
\frac{a}{3}\Phi^3\right) \;,
\end{equation}
where $a$ is an arbitrary constant, one finds the following one-parameter
family of supersymmetric evolution equations:
\begin{equation}\label{a8}
\Phi_t=-\Phi_{xxx}+3\left(\Phi{\cal D}_1{\cal D}_2\Phi \right)_{x}
+\frac{a-1}{2}\left( {\cal D}_1{\cal D}_2\Phi^2\right)_{x}+
3a\Phi^2\Phi_{x} \quad .
\end{equation}
The $N=1$ super KdV equation can be obtained as a proper
reduction of this $N=2$ one.

It was shown that the equation (\ref{a8}) is completely integrable,
i.e. possesses
a Lax pair representation and admits an infinite number of the conserved
quantities, only for $a=-2,\;4$ ${}^{7}$. For $a=1$ there still exist
higher-order conservation laws${}^{8}$, however, no
standard Lax representation is known. Hence,
the proof of complete integrability of the $N=2$ super KdV equation for
$a=1$ is an open problem.

A natural extension of the above scheme to the $N=3$ case we are interested
in is to start with the $N=3$ supercurrent which is subject to the SOPE
relations (or, equivalently, the Poisson brackets) generating
$N=3$ superconformal algebra${}^{11}$,
to construct the appropriate Hamiltonian out of
this supercurrent and to define the $N=3$ super KdV equation as the evolution
equation with respect to this Hamiltonian structure. This is what we do
in Sect.2 of the present paper. We show that the most general $N=3$ super
KdV Hamiltonian (respecting the automorphism $SO(3)$ symmetry along
with $N=3$ supersymmetry), like in the $N=2$ case, involves one free
parameter, thus generating a one-parameter family of the $N=3$ super
KdV equations {\footnote{In${}^{10}$ a particular case
of this general Hamiltonian has been considered, it corresponds to the zero
value of the parameter.}. Requiring the $N=3$ KdV equation to yield,
upon the reduction $N=3 \rightarrow N=2$, one of the integrable (or
``would-be integrable'') versions of the $N=2$ KdV equation fixes
the parameter at some non-zero values. Unfortunately, and this is the radical
difference from the lower $N$ cases, even for these special values of the
parameter the $N=3$ KdV equation turns out to be non-integrable: it does not
admit the Lax representation (at least in the form employed earlier in the
$N=0,\;N=1$ and
$N=2$ cases) and nontrivial local higher order conservation laws.
\setcounter{footnote}0

In Sect.3, in order to clear up the origin of this
difficulty, we analyze the question of existence of the first
non-trivial higher order conservation law for the most general $N=3$ super
KdV equation containing several free parameters.
We find that requiring the existence of such a
conservation law unambiguously fixes all the unknown coefficients in the
$N=3$ super KdV equation.
The resulting equation is different from that constructed in Sect. 2.
Upon the reduction to $N=2$ it turns out to yield
just the special would-be integrable case of the $N=2$ super KdV equation with
$a=1$.
It contains, as its bosonic core,
the coupled system of the ordinary KdV equation
for the dimension 2 scalar field $u(x)$
(conformal stress-tensor)
and the special version of the matrix modified KdV one for the $SO(3)$
triplet of the dimension 1 fields $v^i(x)$
($SO(3)$ Kac-Moody currents).

In Sect.4 we address the problem of the Hamiltonian
description of our new $N=3$ super KdV equation.
We find that it can be
obtained as a closed subsystem of an enlarged system of
the superfield equations involving
an extra $N=3$ superfield $\tilde{J}$. The latter generates
a centrally extended
$N=3$ superconformal algebra while the KdV superfield $J$ itself
is now treated as
{\it quasi-primary} with respect to $\tilde{J}$, with an additional central
charge. On its own right $J$ generates a commutative superalgebra.

\setcounter{equation}{0}
\section{$N=3$ super KdV from $N=3$ superconformal algebra}

For deducing an $N=3$ extension of the KdV equation we can try the same
strategy as in the $N=0$, $N=1$ and $N=2$ cases. Namely, we
choose as the basic object a $N=3$ conformal supercurrent
\be
J(Z) = \psi (z) + \theta^{i}v^{i} (z) +\theta^{3-i}\xi^{i} (z)
+\theta^3 u(z)
\ee
where
$Z=(x,\theta^{i}), i=1,2,3$
are the coordinates of $N=3,\;1D$ superspace,
\be
\theta^3 = \frac{1}{6}\epsilon^{kji}\theta^i \theta^j \theta^k \quad ,
\quad \theta^{3-i}=\frac{1}{2}\epsilon^{kji}\theta^j \theta^k
\ee
and the components $\psi (x),\;v^i (x),\;\xi^i (x),\;u(x)$ form the
supermultiplet of currents of $N=3$ superconformal algebra${}^{11}$
(respectively, the dimension $\frac{1}{2}$ singlet fermionic current, the
triplet of the dimension 1 $SO(3)$ Kac-Moody currents, the triplet
of the dimension $\frac{3}{2}$ fermionic currents and the conformal
stress-tensor
of dimension 2). The structure relations of the $N=3$ superconformal
algebra with an arbitrary central charge $c$ can be summarized as the
following Poisson brackets between
the supercurrents $J(Z),\;J(Z')$ :
\begin{equation}\label{a9}
\left\{  J(Z),J (Z')\right\}_+  =  \left[ \frac{c}{12}{\cal D}^3
 -\frac{1}{2}J \partial +\frac{1}{2}{\cal D}^i J {\cal D}^i
+\partial J\right] \Delta(Z-Z') \quad ,
\end{equation}
where we denoted
$$\Delta (Z-Z')=\frac{1}{6}\epsilon^{ijk}({\theta^i}-{\theta^i}')
({\theta^j}-{\theta^j}')({\theta^k}-{\theta^k}')\delta (z-z')
$$
and defined the spinor covariant derivatives
\begin{equation}\label{a10}
\D{i}=\frac{\partial}{\partial\theta^{i}}
 -\theta^{i}\frac{\partial}{\partial x} \quad , \quad
\left\{\D{i},\D{j} \right\}= -2\delta^{ij}\partial_x \quad ,
\end{equation}
$$
\D{3}=\frac{1}{6}\epsilon^{ijk}\D{i}\D{j}\D{k} \quad , \quad
\D{3-i}=\frac{1}{2}\epsilon^{ijk}\D{j}\D{k} \quad , \quad
\D{3-ij}=\epsilon^{ijk}\D{k} \quad .
$$
The $N=3$ supercurrent $J(Z)$ has the dimension $\frac{1}{2}$, so the
most general Hamiltonian having the dimension 3 (needed for the correspondence
with the bosonic KdV) and respecting both $N=3$ supersymmetry and the
automorphism $SO(3)$ symmetry is given by the expression
\begin{equation}
{\cal H}=\int dxd^{3}\theta \left( J\D{3}J + \frac{\alpha}{3} J\D{i}J\D{i}J
\right) \; ,
\label{a11}
\end{equation}
where $\alpha$ is an arbitrary parameter and the specific normalization
has been chosen for further convenience (c.f. eq. (\ref{a7})).
Using the Poisson structure (\ref{a9})
it is then easy to verify that the Hamilton equation
\begin{equation}\label{a12}
J_t=\left\{ J,\;{\cal H} \right\}
\end{equation}
yields the following two-parameter family of the evolution equations for the
supercurrent $J(Z)$
\begin{equation}\label{a13a}
 J_t=-\frac{c}{6}J_{xxx}+3\left( J\D{3}J \right)_{x}
+\frac{6-c\alpha}{12}\D{3}\left(\D{i}J\D{i}J\right)
+\frac{12-c\alpha}{6}\D{3}\left( J\partial J\right)
+\alpha  \left( J\D{i}J\D{i}J\right)_{x} \quad .
\end{equation}
Note that we are at freedom to fix the central charge $c$
at any non-zero value by rescaling the
variables in eq. (\ref{a13a}) as
$$
t\rightarrow \frac{1}{b}t \quad , \quad
J\rightarrow bJ \quad , \quad
\alpha \rightarrow \frac{1}{b}\alpha \quad ,
$$
$b$ being an arbitrary parameter. So we actually deal with the one-parameter
family. It is convenient to choose $c=6$. Eventually, the $N=3$ super KdV
equation we will discuss in this Section is as follows
\begin{equation}\label{a13}
 J_t=-J_{xxx}+3\left( J\D{3}J \right)_{x}
+\frac{1-\alpha}{2}\D{3}\left(\D{i}J\D{i}J\right)
+(2-\alpha)\D{3}\left( J\partial J\right)
+\alpha  \left( J\D{i}J\D{i}J\right)_{x} \quad .
\end{equation}

It remains to find out whether the parameter $\alpha$ can be chosen so that
the associated equation is completely integrable as in the $N=2$ case, i.e.
admits a Lax pair representation and exhibits infinitely many conservation
laws.

To start with, we note that eq.(\ref{a13}) has a proper reduction
to the $N=2$ case. Indeed, if we choose
\begin{equation}\label{a14}
J(Z)=\theta^{(3)}\Phi \quad ,
\end{equation}
then we immediately obtain the $N=2$ super KdV equation (\ref{a8})
with
$$
a=\alpha \quad .
$$
It is clear that the integrable version of the
$N=3$ super KdV equation (if exists) should
yield the integrable $N=2$ super KdV upon the reduction. So it is
natural to limit our study to the following values
of $\alpha$:
$$
\alpha_1= -2\; , \;\; \alpha_2=4\;, \;\; \alpha_3=1 \;,
$$
which correspond, respectively, to the two integrable and one would-be
integrable $N=2$ super KdV equations.

Unfortunately, our equation (\ref{a13}) admits no
standard Lax representation in the form ${}^{8}$
$$ L_t = [-4\; L^{3\ /2}_{+}, L ]$$
for any value of $\alpha$. We have checked this by a tedious but
straightforward computation. One might think that, like in the $N=2$ case,
eq. (\ref{a13}) could have higher order conservation laws despite
the non-existence of Lax representation. However, our attempts to find
non-trivial higher order
conservation laws reducible to those of the $N=2$ super KdV upon the
reduction $N=3 \rightarrow N=2$ have
also failed for any value of $\alpha$. Thus a straightforward application of
the approach used
previously for constructing integrable KdV equations in the $N=0$, $N=1$
and $N=2$ cases leads to a non-integrable system in the $N=3$ case.
In the next Section we propose another way to obtain an
integrable $N=3$ super KdV equation by
considering the most general $N=3$
superfield extension of the KdV equation and finding the conditions under which
it possesses non-trivial higher order conservation laws.

\setcounter{equation}{0}
\section{N=3 super KdV and conservation laws.}

Now we turn to an explicit construction of $N=3$ supersymmetric KdV
equation possessing non-trivial conservation laws. We postpone to Sect.4
a discussion of how it can be given a Hamiltonian interpretation.

Under natural conditions of $N=3$ supersymmetry and $SO(3)$ symmetry
the most general $N=3$ super KdV equation is of the form
\be \label{c1}
J_{t}={\cal A}(J) \quad ,
\ee
where ${\cal A}$ is a linear combination of all possible terms with proper
dimension (7/2)
which can be constructed from the $N=3$ superfield $J(Z)$ and covariant
spinor derivatives.
Explicitly, it is the six-parameter family of equations
\bea
J_{t}& = & -J_{xxx} +a_{1} \left( J\D{3} J\right)_{x}
+a_{2} \D{3}\left( JJ_{x}\right) +a_{3}\D{3} \left( \D{i}J\D{i}J\right)
\nonumber \\
 & & +a_{4}\left( \D{i}J\right)_{x}\D{3-i}J
+a_{5}J\left( \D{i}J\D{i}J\right)_{x} +a_{6}J_{x}\left( \D{i}J\D{i}J\right)
\quad . \label{c2}
\eea
In order to reduce the number of parameters and thereby to simplify
computations we impose the requirement
that upon the reduction to the $N=2$ case eq. (\ref{c2}) goes over to the
known $N=2$ KdV family (\ref{a8}). This condition gives rise to the following
relations between the parameters $a_{1},\ldots ,a_{6}$:
\be
a_{1}=3 \quad , \quad
a_{3}=\frac{1-a}{2} \quad , \quad
a_{4}=0 \quad , \quad
2a_{5}+a_{6}=3a \quad ,
\ee
where $a$ is the parameter which enters the $N=2$ super KdV equation.
So, the $N=3$ super KdV equation we will consider
contains three undetermined parameters
\bea
J_{t}& = & -J_{xxx} +3 \left( J\D{3} J\right)_{x}
+a_{2} \D{3}\left( JJ_{x}\right)
+\frac{1-a}{2}\; \D{3} \left( \D{i}J\D{i}J\right)
\nonumber \\
 & & +\frac{1}{2}(3a-a_{6})\; J\left( \D{i}J\D{i}J\right)_{x}
+a_{6}J_{x}\left( \D{i}J\D{i}J\right)
\quad . \label{c3}
\eea
The previously considered equation (\ref{a13}) is the particular case of
(\ref{c3}) corresponding to the choice
$$a_2 = 2- a \;, \;\;\; a_6 = a \;.$$

Now we wish to inquire whether this three-parameter
family of equations yields integrable systems for some
specific values of the parameters. Here we do not concern the
question of the existence of the relevant Lax pairs. Instead we search for
the first non-trivial higher order conservation law.

The simplest candidate for the higher order conserved
quantity is an integral of degree 5
over $N=3$ superspace with the integrand constructed from all possible
independent densities of degree 9/2,
each multiplied by an undetermined coefficient{\footnote{Recall that
the $N=3$ superspace integration measure $(dx d\theta^3)$
has the dimension 1/2, so the integral (\ref{c4}) has the dimension 5.}
\bea
H_5 & = & \int dxd^{3}\theta \{
A_{1}J\D{3}J_{xx}+A_{2}J\D{i}J\D{i}J_{xx}+A_{3}JJ_{x}J_{xx}+
   A_{4}J\D{3}J\D{3}J  \nonumber \\
 & &  + A_{5}JJ_{x}\D{i}J\D{3-i}J + A_{6}J\D{i}J\D{i}J\D{3}J+
 J\left( \D{i}J\D{i}J \right)^2 \}   \quad . \label{c4}
\eea
The coefficients are then fixed by requiring the integral to be conserved
(i.e. time-independent) on the equation of motion (\ref{c3}),
$$ \left( H_5 \right)_t = 0\;.$$
This also must fix the values of parameters $a,\;a_2,\;a_6$
in (\ref{c3}).
\setcounter{footnote}0

After tedious calculations one finds that {\it all coefficients}
in the integral (\ref{c4}) and in eq. (\ref{c3}) are fixed to the
unique values
\be
A_1=-5\;, \;\; A_2=-\frac{5}{2}\;, \;\; A_3=\frac{5}{2}\;, \;\;
A_4=10\;,\;\;A_5=\frac{5}{3}\;, \;\;
A_6=\frac{20}{3} \label{c5}
\ee
\be
a=1 \;,\;\; a_2 =0\;, \;\; a_6=0 \quad .
\label{c6}
\ee
Thus in the $N=3$ supersymmetric case there exists only one superfield
extension of the KdV equation which possesses a nontrivial higher order
conservation law
\be
J_{t} =  -J_{xxx} +3 \left( J\D{3} J\right)_{x}
 +\frac{3}{2} J\left( \D{i}J\D{i}J\right)_{x}
\quad . \label{c7}
\ee
It is curious that after reduction to the $N=2$ case this equation
goes over to the exceptional $N=2$ super KdV equation with parameter $a=1$.
For completeness we write also the first two lower order conserved
quantities for eq. (\ref{c7})
\bea
H_1 & = & \int dx d^{3}\theta J  \nonumber \\
H_3 & = & \int dx d^{3} \theta \left(  J \D{3} J
+\frac{1}{3} J\D{i} J\D{i} J \right)  \label{c8}
\eea

A few comments are needed concerning the equation (\ref{c7}).

First of all,
we have started with the most general $N=3$ superfield equation (\ref{c2}).
The only extra demand we have employed from the beginning
was the existence of a proper
reduction to the $N=2$ case. It seems very intriguing that under such
general assumptions we were eventually left with the unique
candidate
for the integrable $N=3$ KdV equation.

Secondly, recall that even for the $N=2$ super KdV equation
the integrability at $a=1$ is an open problem due to lacking
of the standard Lax representation in this case. The problem
of proving integrability remains, of course, in our case too. Up to now we
know only the first non-trivial conservation law for the equation (\ref{c7}).
Let us stress, however, that the
set of equations that must be satisfied by the coefficients $a,\;a_{i},\;A_{i}$
is highly overdetermined. There are about five times as many equations
compared to the unknowns.
So the very existence of this first nontrivial conservation
law is a strong indication for the complete integrability of the
corresponding equation.

Finally, we briefly discuss the bosonic core of our $N=3$ super KdV
equation (\ref{c7}).

It is straightforward to find the set of bosonic equations to which
eq. (\ref{c7}) is reduced after putting all fermions equal to zero
\begin{eqnarray}
 u_t&=& -u_{xxx}+ 3 \left( u^2-v^i v^i_{xx} + u v^{i}v^{i} \right)_{x}
\nonumber \\
v^i_t&=&-v^i_{xxx} + 3 \left( uv^i \right)_{x}+
   3v^i v^{j}v^{j}_{x} \;, \label{c9}
\end{eqnarray}
where
$$
v^i=\D{i}J\vert \quad , \quad u={\cal D}^3J\vert \quad .
$$
It is a crucial novel feature of the $N=3$ KdV equation compared to
the $N=2$ one that
in its bosonic sector, besides the dimension 2 KdV field $u(x)$ which
is identified with
a conformal stress-tensor and generates a Virasoro subalgebra in the $N=3$
superconformal algebra (\ref{a9}), there is also a triplet of the
dimension 1 fields $v^{i}(x)$ which
generate an $SO(3)$ Kac-Moody subalgebra of (\ref{a9}). In the $N=2$ case
only one such a field is present and it generates a $U(1)$ Kac-Moody algebra.

So we see
that the bosonic subsector of our $N=3$ super KdV equation contains
the two coupled equations -- the KdV
equation for the scalar field $u$ and a three-component
generalization of the mKdV equation, both with the extra mixed terms
in the r.h.s. These equations cannot be decoupled
by a redefinition of $u$. While the first equation is a kind of
the perturbed KdV equation, the second one can be viewed as a perturbation of
the equation
\be \label{c10}
v^i_t\;=\;-v^i_{xxx} +
   3v^i ( v^{2} )_{x} \;,
\ee
which is a particular case of the general $SO(3)$ matrix mKdV equation
\begin{equation}\label{c11}
v_t=-v_{xxx} + A \frac{i}{2} [ v, v_{xx}] +B v_x \;(v^2) +Cv\; (v^2)_x\;\;,
\;\;\; v \equiv v^i \tau^i\;,
\end{equation}
$\tau^i$ being Pauli matrices and $A,\;B,\;C$ arbitrary numerical
coefficients. Eq. (3.11) arises under the choice
\be  \label{c12}
A=B=0\;,\;\;\;C=\frac{3}{2}\;.
\ee
Note that in ref.${}^{12}$ the integrability has been shown for another
particular case of eq.(\ref{c11}) corresponding to the option
$$
A=1\;, \;\;B=-C=\frac{1}{6}\;.
$$
Our consideration suggests that, being extended to a coupled system including
a KdV-type equation, this matrix mKdV equation can be as well integrable
for the choice of parameters as in eq.(\ref{c12}).
Anyway, it is clear that the complete
analysis of the
integrability properties of the new $N=3$ super KdV equation (\ref{c7})
should essentially rely upon the study of such properties of the bosonic
subsystem (\ref{c9}) and the matrix mKdV equation (\ref{c11}). We hope to
return to these issues in the future.

\setcounter{equation}{0}
\section{The Hamiltonian structure of new N=3 super KdV equation}

In the previous Section we have found the unique $N=3$ super KdV equation
(\ref{c7}) which possesses a nontrivial higher-order conserved quantity. This
equation cannot be obtained within the standard Hamiltonian approach of
Section 2 as a Hamiltonian flow on $N=3$ superconformal algebra. Indeed, the
only conserved quantity having the dimension of the
Hamiltonian for eq.(\ref{c7}) is $H_3$ defined in eq. (\ref{c8}). It is easy
to see that it coincides with the Hamiltonian (\ref{a11}) at $\alpha =1$.
So the equation produced for $J$ by this Hamiltonian via
the Poisson structure (\ref{a9}) is a particular case of eq.(\ref{a13a}). But
this is just the non-integrable case we started with.

Thus in order to give a Hamiltonian interpretation to $N=3$ super KdV
equation (\ref{c7}) we need to examine the question of existence of another
Hamiltonian structure for this system.

The only way to construct a Hamiltonian formalism for
eq.(\ref{c7}) we have succeeded to invent is to introduce one more spinor
$N=3$
superfield ${\tilde J}$ and to re-obtain (\ref{c7}) as a closed subsystem of
some Hamiltonian system of equations for this extended set of superfields.

Let us start from two independent $N=3$ supercurrents $J_{1}(Z)$ and
$J_{2}(Z)$ and assume that the Poisson bracket structure for these
superfields is given by a direct product of the two standard structures
(\ref{a9}) with arbitrary central charges $c_1$ and $c_2$:
\begin{eqnarray}
\left\{ J_{1}(Z),\;J_{2}(Z')\right\}_+ & = & 0 \nonumber \\
\left\{  J_{1}(Z),\;J_{1}(Z')\right\}_+  & = & \left[ \frac{c_{1}}{12}{\cal
D}^3
 -\frac{1}{2}J_{1} \partial +\frac{1}{2}{\cal D}^i J_{1} {\cal D}^i
+\partial J_{1}\right] \Delta(Z-Z')    \label{d1} \\
\left\{  J_{2}(Z),\;J_{2} (Z')\right\}_+  & = & \left[
\frac{c_{2}}{12}{\cal D}^3
 -\frac{1}{2}J_{2} \partial +\frac{1}{2}{\cal D}^i J_{2} {\cal D}^i
+\partial J_{2}\right] \Delta(Z-Z')    \nonumber
\end{eqnarray}

In other words, at this step we deal with two independent $N=3$
superconformal algebras,
$J_1$ and $J_2$ being the relevant supercurrents.

Now we wish to show that the second Hamiltonian structure for eq. (\ref{c7})
can be obtained as a contraction of the product structure (\ref{d1}).
To this end, let us
pass to the new superfields $J$ and ${\tilde J}$ defined as follows
\begin{equation}\label{d2}
J=J_1-J_2 \quad , \quad {\tilde J}=J_1+J_2 \quad .
\end{equation}
These objects, respectively $\tilde{J}$ and $J$, can be identified with
the supercurrents generating the
diagonal $N=3$ superconformal group in the above product and the coset
over this subgroup. The Poisson bracket structure for these new
superfields is simply another form of
(\ref{d1})
\begin{eqnarray}
\left\{ J(Z), \;J(Z')\right\}_+ & = &
\left[ \frac{c_{1}+c_{2}}{12}{\cal D}^3
 -\frac{1}{2}\tilde{J} \partial +\frac{1}{2}{\cal D}^i \tilde{J} {\cal D}^i
+\partial \tilde{J}\right] \Delta(Z-Z')  \nonumber \\
\left\{ \tilde{J}(Z),\;J(Z')\right\}_+  & = &
\left[ \frac{c_{1}-c_{2}}{12}{\cal D}^3
 -\frac{1}{2}J \partial +\frac{1}{2}{\cal D}^i J {\cal D}^i
+\partial J \right] \Delta(Z-Z')    \label{d3} \\
\left\{ \tilde{J}(Z),\;\tilde{J} (Z')\right\}_+  & = &
\left[ \frac{c_{1}+c_{2}}{12}{\cal D}^3
 -\frac{1}{2}\tilde{J} \partial +\frac{1}{2}{\cal D}^i \tilde{J} {\cal D}^i
+\partial \tilde{J}\right] \Delta(Z-Z')  \quad .  \nonumber
\end{eqnarray}

Let us now deform this structure in the following self-consistent way:
\begin{eqnarray}
J & \rightarrow & \frac{1}{\kappa}J \;,\;\; (c_1-c_2)\equiv
\frac{1}{\kappa}\;c\;,\;\; (c_1 +c_2) \equiv \tilde{c} \label{d4} \\
 \kappa &\rightarrow & 0 \; .  \nonumber
\end{eqnarray}
In the contraction limit (\ref{d4}) goes over to
\begin{eqnarray}
\left\{ J(Z),\;J(Z')\right\}_+ & = & 0  \nonumber \\
\left\{ \tilde{J}(Z),\;J(Z')\right\}_+  & = &
\left[ \frac{c}{12}{\cal D}^3
 -\frac{1}{2}J \partial +\frac{1}{2}{\cal D}^i J {\cal D}^i
+\partial J \right] \Delta(Z-Z')    \label{d5} \\
\left\{ \tilde{J}(Z),\;\tilde{J} (Z')\right\}_+  & = &
\left[ \frac{\tilde{c}}{12}{\cal D}^3
 -\frac{1}{2}\tilde{J} \partial +\frac{1}{2}{\cal D}^i \tilde{J} {\cal D}^i
+\partial \tilde{J}\right] \Delta(Z-Z')  \quad .  \nonumber
\end{eqnarray}

Now we consider the most general $N=3$ supersymmetric (and $SO(3)$ symmetric)
Hamiltonian which is linear in
${\tilde J}$
\begin{equation}\label{d6}
H=\int dx d^3 \theta \left( \gamma {\tilde J}\D{3} J
+\alpha {\tilde J}\D{i} J \D{i} J +
\beta {\tilde J} J J_{x} \right) \quad .
\end{equation}
This Hamiltonian gives rise to the following evolution equation:
\begin{eqnarray}
J_t & = & -\frac{c\gamma}{12}J_{xxx} + \frac{3\gamma}{2} \left(
J\D{3} J\right)_{x}+\left(\frac{\gamma}{4}+\frac{c\alpha }{12}\right)
\D{3}\left(\D{i} J\D{i} J\right) + \left( \gamma +\frac{c\beta}{12}\right)
  \D{3}\left( J J_{x} \right) \nonumber \\
   & & - \frac{2\alpha + \beta}{4}J \left( \D{i}J\D{i}J\right)_{x}
  -\frac{4\alpha - \beta }{4} J_{x} \D{i}J\D{i}J \quad . \label{d7}
\end{eqnarray}
Making in eq.(\ref{d7}) arbitrary rescalings of $x$, $t$, $\theta$ and $J$,
and observing that only two
of these rescalings are actually independent, we are at liberty to fix
two parameters. We choose the following option
\begin{equation}\label{d8}
 \gamma =2 \quad , \quad c=6  \quad .
\end{equation}
As a result our equation takes the form
\begin{eqnarray}
J_t & = & -J_{xxx} + 3 \left( J\D{3} J\right)_{x}+\frac{\alpha +1}{2}
\D{3}\left(\D{i} J\D{i} J\right) + \frac{\beta +4}{2}
  \D{3}\left( J J_{x} \right) \nonumber \\
   & & - \frac{2\alpha + \beta}{4}J \left( \D{i}J\D{i}J\right)_{x}
  -\frac{4\alpha - \beta }{2} J_{x} \D{i}J\D{i}J\;\;. \label{d9}
\end{eqnarray}

If we now compare this equation with our previous equation (\ref{c7}),
we immediately find that they coincide for the following values of
parameters $\alpha$ and $\beta$:
\begin{equation}
\alpha = -1 \quad , \quad \beta = -4 \quad .
\end{equation}
For $\tilde J$ one also obtains some evolution equation whose
precise form is of no interest for us here.

So we have succeeded in interpreting our $N=3$ super KdV equation
as a Hamiltonian equation in
the framework of an extended system which includes the additional superfield
${\tilde J}$. It is worthwhile to emphasize that in this approach the
KdV superfield $J$ generates a commutative translation superalgebra
instead of the $N=3$ superconformal algebra; the crucial point in deducing
eq.(\ref{c7}) from the
Hamiltonian (\ref{d6}) is that $J$
behaves as a "quasi-primary superfield" with
respect to an extra $N=3$ superconformal algebra generated by $\tilde{J}$.
This latter property manifests itself as the presence of a nonvanishing
central charge $c$ in the second relation (\ref{d5}).

It is worth mentioning that the scalar field KdV equation (\ref{a1})
can also be obtained starting from the system of two scalar
fields $u(x), {\tilde u}(x)$ with the Poisson bracket structure given by
\begin{eqnarray}
\left\{ u(x),\;u(y) \right\} & = & 0 \nonumber \\
\left\{ {\tilde u}(x),\;u(y) \right\} & = &
\left[ -\partial^3+4u\partial+2u_x
  \right]\delta (x-y) \label{d10} \\
\left\{ {\tilde u}(x),\;{\tilde u}(y) \right\} & = &
\left[ -\partial^3+4{\tilde u}\partial+2{\tilde u}_x
  \right]\delta (x-y) \nonumber
\end{eqnarray}
and the Hamiltonian
\begin{equation}\label{d11}
H=\frac{1}{2}\int  dx {\tilde u}u  \quad .
\end{equation}
This doubling of fields looks rather artificial for the scalar KdV equation,
owing to the existence of the standard
Hamiltonian (\ref{a2}), but the lacking of such a Hamiltonian for the
$N=3$ super KdV equation (\ref{c7}) immediately leads us to make use of
this possibility (it is the only one known to us at present).

Let us note, at the end of this Section, that almost all known systems
with $N=3$ supersymmetry respect as well $N=4$ supersymmetry. Thus,
the above doubling of fields could perhaps be interpreted as an extension of
our $N=3$ multiplet of currents to the $N=4$ one or at least as coming
from a contraction of the second Hamiltonian structure
for $N=4$ super KdV equation. This question certainly
warrants further investigation. We postpone its discussion to the future.

\section{Conclusion}

In this paper we have
demonstrated that in the case of the $N=3$ super KdV equation the standard
second Hamiltonian structure based on $N=3$ superconformal algebra gives rise
to
a non-integrable system. We have
deduced a new $N=3$ super KdV equation by considering the most general
$N=3$ superextension of the KdV equation and checking the existence of
the higher order non-trivial superfield conservation laws for it. It is
interesting
that there exists a unique $N=3$ superextension of the KdV equation which
possesses non-trivial conservation laws. After reduction to the $N=2$
case this equation turns into the exceptional $N=2$ super KdV equation
(with parameter $a=1$) the integrability of which is under
investigation${}^{8}$.
The bosonic core of our modified $N=3$ super KdV equation
contains the new system of coupled KdV and matrix mKdV equations which has a
great chance to be integrable.

We have also proposed the Hamiltonian structure for our $N=3$ super
KdV equation. It appears
as some contraction of the direct sum of two $N=3$ superconformal
algebras. It is an open question whether this structure can be
somehow related to $N=4$ superconformal algebras. So it seems very
interesting to consider possible integrable $N=4$ superextensions of the
KdV equation.

\setcounter{equation}{0}
\def\thesection { }
\section{Acknowledgements}
We are grateful to I. Batalin, J. Lukierski and Z. Popowicz for many
useful and clarifying discussions.

\end{document}